\documentclass[twocolumn,prl,superscriptaddress]{revtex4}
\usepackage[latin9]{inputenc}
\usepackage{amsmath}
\usepackage{amssymb}
\usepackage{graphicx}
\usepackage{esint}

\makeatletter
\@ifundefined{textcolor}{}
{%
 \definecolor{BLACK}{gray}{0}
 \definecolor{WHITE}{gray}{1}
 \definecolor{RED}{rgb}{1,0,0}
 \definecolor{GREEN}{rgb}{0,1,0}
 \definecolor{BLUE}{rgb}{0,0,1}
 \definecolor{CYAN}{cmyk}{1,0,0,0}
 \definecolor{MAGENTA}{cmyk}{0,1,0,0}
 \definecolor{YELLOW}{cmyk}{0,0,1,0}
}

\def\@dotsep{4.5}

\usepackage{bm}
\usepackage{amsbsy}\usepackage{amsbsy}\usepackage{epsf}\usepackage{graphics}\usepackage{rotating}\usepackage{bm}\usepackage{enumerate}\usepackage{color}

\makeatother

\begin{document}

\title{A Path Factorization Approach to Stochastic Simulations}

\author{Manuel Ath\`enes}

\affiliation{CEA, DEN, Service de Recherches de M\'etallurgie Physique, F-91191
Gif-sur-Yvette, France}

\author{Vasily V. Bulatov}

\affiliation{Lawrence Livermore National Laboratory, Livermore, California 94551,
USA}






\begin{abstract}
The computational efficiency of stochastic simulation algorithms is
notoriously limited by the kinetic trapping of the simulated trajectories
within low energy basins. Here we present a new method that overcomes
kinetic trapping while still preserving exact statistics of escape
paths from the trapping basins. The method is based on path factorization
of the evolution operator and requires no prior knowledge of the underlying
energy landscape. The efficiency of the new method is demonstrated
in simulations of anomalous diffusion and phase separation in a binary
alloy, two stochastic models presenting severe kinetic trapping. 
\end{abstract}

\maketitle

Time evolution of many natural and engineering systems is described
by a master equation (ME), i.e. a set of ordinary differential equations
for the time-dependent vector of state probabilities~\cite{vankampen:2007,redner:2001}.
For models with large (or infinite but countable) number of states,
direct solution of the ME is prohibitive and kinetic Monte Carlo (kMC)
is used instead to simulate the time evolution by generating sequences
of stochastic transitions from one state to the next~\cite{lanore:1974,BKL,gillespie:1977}.
Statistically equivalent to the (most often unknown) solution of the
ME, kMC finds growing number of applications in natural and engineering
sciences. However still wider applicability of kMC is severely limited
by the notorious kinetic trapping where the stochastic trajectory
repeatedly visits a subset of states, a trapping basin, connected
to each other by high-rate transitions while transitions out of the
trapping basin are infrequent and take great many kMC steps to observe.

In this Letter, we present an efficient method for sampling stochastic
trajectories escaping from the trapping basins. Unlike recent methods
that focus on short portions of the full kinetic path directly leading
to the escapes and/or require equilibration over a path ensemble~\cite{dellago:2002,sun:2006,harland:2007,vansiclen:2007,eidelson:2012,mora:2012,manhart:2013},
our method constructs an entire stochastic trajectory within the trapping
basin including the typically large numbers of repeated visits to
each trapping state as well as the eventual escape. Referred hereafter
as kinetic Path Sampling (kPS), the new algorithm is statistically
equivalent to the standard kMC simulation and entails (i) iterative
factorization of paths inside a trapping basin, (ii) sampling a single
exit state within the basin's perimeter and (iii) generating a first-passage
path and an exit time to the selected perimeter state through an exact
randomization procedure. We demonstrate the accuracy and efficiency
of kPS on two models: (1) diffusion on a random energy landscape specifically
designed to yield a wide and continuous spectrum of time scales and
(2) kinetics of phase separation in super-saturated solid solutions
of copper in iron. The proposed method is immune to kinetic trapping
and performs well under simulation conditions where the standard kMC
simulations slows down to a crawl. In particular, it reaches later
stages of phase separation in the Fe-Cu system and captures a qualitatively
new kinetics and mechanism of copper precipitation.

The evolution operator, obtained formally from solutions of the ME, can be 
expressed as an exponential of the time-independent transition rate matrix~\cite{footnote:me} 
\begin{equation}
{\mathbf{P}}(t,t+\tau)=\exp\left(\int_{t}^{t+\tau}\mathbf{M}ds\right)=\exp\left(\tau\mathbf{M}\right),\label{eq:propagator}
\end{equation}
where $P_{\beta\gamma}(t,t+\tau)$ is the probability to find the system in state $\gamma$ at $t+\tau$
given that it was in state $\beta$ at time $t$, $M_{\beta\gamma}$
is the rate of transitions from state $\beta$ to state $\gamma$
(off-diagonal elements only) and the standard convention is used to
define the diagonal elements as $M_{\beta\beta}=-\sum_{\nu \neq \beta}M_{\beta\nu}$.
As defined, the evolution operator belongs to the class of stochastic
matrices such that $\sum_{\nu}P_{\beta\nu}=1$ and $P_{\beta\gamma}\ge0$
for any $\beta$, $\gamma$, $t$ and $\tau$. If known, the evolution
operator can be used to sample transitions between any two states
and over arbitrary time intervals $\tau$~\cite{note:transition}.
In particular, substantial simulation speed-ups can be achieved by
sampling transitions to distant states on an absorbing perimeter of
a trapping basin. Two main defficiencies of the existing implementations
of this idea ~\cite{novotny:1995,boulougouris:2007,barrio:2013,nandipati:2010}
is that states within the trapping basin are expected to be known
{\em a priori} and that computing the evolution operator requires
a partial eigenvalue decomposition of $\mathbf{M}$ entailing high
computational cost ~\cite{moler:2003}. In contrast, the kPS algorithm
does not require any advance knowledge of the trapping basin nor does
it entail matrix diagonalization. Instead, kPS detects kinetic trapping
and charts the trapping basin iteratively, state by state, and achieves
high computational efficiency by sequentially eliminating all the
trapping states through path factorization~~\cite{athenes:1997,trygubenko:2006,wales:2009}.
Here, Wales's formulation~\cite{wales:2009} of path factorization
is adopted for its clarity.

Consider the linearized evolution operator 
\begin{equation}
\mathbf{P}^{(0)}(t,t+\tau)=\mathbf{I}+\tau\mathbf{M},\label{eq:linear}
\end{equation}
where $\mathbf{I}=\mathbf{P}^{(0)}(t,t)$ is the identity matrix.
Assuming that $\tau \le \min \left\{ -(M_{\beta\beta})^{-1} : \forall \beta \right\} $, $\mathbf{P}^{(0)}$ is a proper stochastic matrix
that can be used to generate stochastic sequences of states from the
ensemble of paths defined by matrix $\mathbf{M}$. The diagonal elements
of $\mathbf{P}^{(0)}$ define the probabilities of round-trip transitions
after which the system remains in the same state. To correct for the
linearization of the evolution operator in~\eqref{eq:linear}, the
time elapsed before any transition takes place is regarded as a stochastic
variable and sampled from an exponential distribution $t\rightarrow\tau^{-1} \exp(-t/\tau)$
~\cite{serebrinsky:2011}. This simple time randomization obviates
the need for exponentiating the transition rate matrix in~\eqref{eq:propagator}.
Following \cite{wales:2009}, consider a bi-directional connectivity
graph defined by $\mathbf{P}^{(0)}$ in which $N$ states in the trapping
basin are numbered in order of their elimination, $\mathbb{E}=\{1,2,...,N\}$.
An iterative path factorization procedure then constructs a set of
stochastic matrices $\{\mathbf{P}^{(n)}\}_{0\leq n\leq N}$ such that,
after the $n$-th iteration, all states $\gamma\leq n$ are eliminated
in the sense that the probability of a transition from any state $\beta$
to state $\gamma\leq n$ is zero. Specifically, at $n$-step of factorization
the transition probability $P_{\beta\gamma}^{(n)}$ ($\gamma>n$,
$\forall\beta$) is computed as the sum of the probability of a direct
transition $P_{\beta\gamma}^{(n-1)}$ and the probabilities of all
possible indirect paths involving round-trips in $n$ after having initially transitioned from $\beta$ to $n$ and before finally transitioning from $n$ to $\gamma$, e.g.  $\beta\rightarrow\gamma$, $\beta\rightarrow n\rightarrow\gamma$,
$\beta\rightarrow n\rightarrow n\rightarrow\gamma$, $\beta\rightarrow n\rightarrow n\rightarrow n\rightarrow\gamma$,
and so on. With the round-trip probability being $P_{nn}^{(n-1)}$,
it is a simple matter to sum the geometric series corresponding to
the round-trip paths. Although any intermediate $\mathbf{P}^{(n)}$
can be used to generate stochastic escapes from any state $\alpha\in\mathbb{E}$,
a trajectory generated using $\mathbf{P}^{(N)}$ is the simplest containing
a single transition from $\alpha$ that effectively subsumes all possible
transitions involving the deleted states in the trapping basin $\mathbb{E}$.
On the other end, a detailed escape trajectory can be generated using
$\mathbf{P}^{(0)}$ that accounts for all transitions within $\mathbb{E}$,
reverting back to the standard kMC simulation. Remarkably, it is possible
to construct a detailed escape trajectory statistically equivalent
to the standard kMC without ever performing a detailed (and inefficient)
kMC simulation. 

Consider matrix $\mathbf{H}^{(n)}$ whose elements $H_{\beta\gamma}^{(n)}$
store the number of $\beta\rightarrow\gamma$ transitions observed
in a stochastic simulation with $\mathbf{P}^{(n)}$. Given $\mathbf{H}^{(n)}$,
one can randomly generate $\mathbf{H}^{(n-1)}$, the matrix similarly
used to count the transition numbers observed in a stochastic process
based on $\mathbf{P}^{(n-1)}$ without actually performing the simulation
using $\mathbf{P}^{(n-1)}$. The ratio of transition probabilities
($\gamma>n$) 
\begin{equation}
R_{\beta\gamma}^{(n)}={P_{\beta\gamma}^{(n-1)}}\slash{{P}_{\beta\gamma}^{(n)}}\label{eq:ratio}
\end{equation}
defines the conditional probability that a trajectory generated using
$\mathbf{P}^{(n-1)}$ contains a direct transition from $\beta$ to
$\gamma$ given that the trajectory generated with $\mathbf{P}^{(n)}$
contains the same transition. For $\beta=n$, $R_{n\gamma}^{(n)}$
is independent of $\gamma$ and is equal to $1-P_{nn}^{(n-1)}$, the
probability of escape from $n$. It is thus possible to generate $\mathbf{H}^{(n-1)}$
by performing a stochastic simulation with $\mathbf{P}^{(n)}$, harvesting
$\mathbf{H}^{(n)}$ and drawing random variates from (standard and negative) binomial distributions whose exponents and coefficients are given
by the elements of $\mathbf{H}^{(n)}$ and $\mathbf{R}^{(n)}$, respectively.
This randomization procedure can be used iteratively on $n$ in the
reverse order from $N$ to $1$ to generate $\mathbf{H}^{(0)}$ containing
a detailed count of transitions involving all states in $\mathbb{E}$.
Finally, the time of exit out of $\mathbb{E}$ is sampled by drawing
a random variate from the gamma distribution whose 
scale and shape parameters are defined by $\tau$ and the total number of transitions contained in $\mathbf{H}^{(0)}$, 
respectively. Thus, in its simplest form the kPS algorithm proceeds
by first deleting all states in $\mathbb{E}$ through iterative forward
path factorization, then using $\mathbf{P}^{(N)}$ to sample a single
transition from $\alpha\in\mathbb{E}$ and to generate $\mathbf{H}^{(N)}$
followed by a backward randomization to reconstruct a detailed stochastic
path within $\mathbb{E}$ and to sample an escape time to the selected
exit state. A detailed description of the kPS algorithm is given in
the Supplemental Material~\cite{athenes:2014}.

\begin{figure}[!h]
\includegraphics[width=0.6\textwidth]{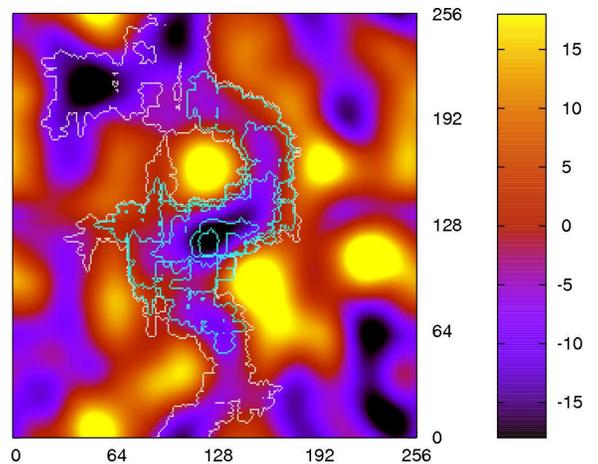} \protect\caption{Energy surface of the saddle points. The color scale is in the units
of $\epsilon$. Artificial smoothing is used for better visualization.}

\label{fig:landscape} 
\end{figure}

We first apply kPS to simulations of a random walker on a disordered
energy landscape (substrate)~\cite{limoge:1990}. The substrate is
a periodically replicated 256$\times$256 fragment of the square lattice
on which the walker hops to its four nearest-neigbour (NN) sites with
transition rates 
\[
M_{\beta\gamma}=\exp\left[(E_{\beta}-E_{\beta\gamma}^{\mathrm{s}})/T\right],
\]
where $\mathrm{T}$ is the temperature, $E_{\beta}$ the site energy
and $E_{\beta\gamma}^{\mathrm{s}}$ the saddle energy between sites
$\beta$ and $\gamma$. The energy landscape is purposefully constructed
to contain trapping basins of widely distributed sizes and depths
(see the Supplemental Material~\cite{athenes:2014} for details)
and is centered around the walker's initial position next to the lowest
energy saddle (Fig.~\ref{fig:landscape}).

When performed at temperature T$=2.5$, standard kMC simulations (with
hops only to the NN sites) are efficient enabling the walker to explore
the entire substrate. However at T$=1$, the walker remains trapped
near its initial position repeatedly visiting states within a trapping
basin. To chart a basin set $\mathbb{E}$ for subsequent kPS simulations,
the initial state 1 is eliminated at the very first iteration followed
in sequence by the ``most absorbing states'' for
which $P_{1\gamma}^{(n)}$ is found to be largest at the $n$-th iteration
($2\le n\le N$). The expanding contours shown in Fig.~\ref{fig:landscape}
depict the absorbing boundary $\partial\mathbb{A}$ (perimeter of
the basin) obtained after eliminating $2^{7}$, $2^{9}$, $2^{11}$,
$2^{12}$, $2^{13}$ and $2^{14}$ states. The perimeter contour $\partial\mathbb{A}$
consists of all states $\gamma$ for which $P_{1\gamma}^{(N)}$ is
nonzero. 
\begin{figure}[!h]
\includegraphics[width=0.45\textwidth]{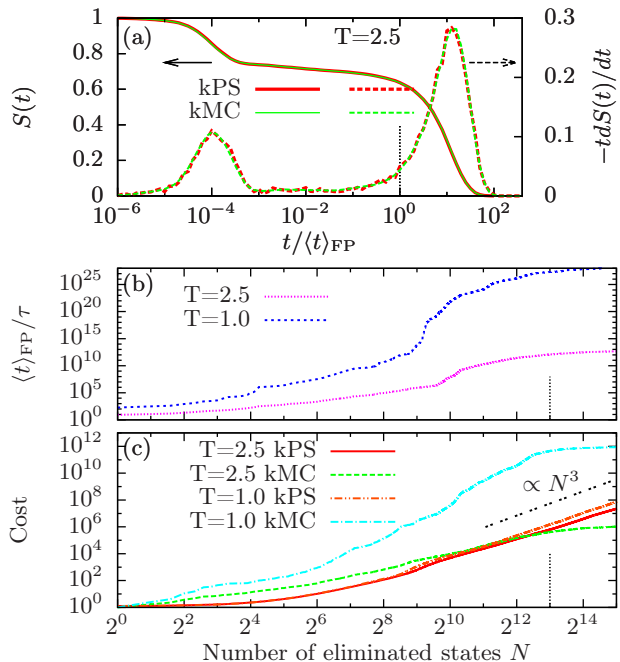} \protect\caption{(a) The probability $S(t)$ for the walker to remain within a trapping
basin containing $N=2^{13}$ states (solid line) and the distribution
of times of escape out of the same basin (dashed lines) using a log scale for the bins; (b) The mean
first-passage time $\langle t\rangle_{\mathrm{FP}}$ as a function
of the number of states $N$ included in the trapping basin; (c) Computational
cost of kMC and kPS simulations as a function of $N$ at two different
temperatures (in the units of a single kMC hop). }

\label{fig2} 
\end{figure}

To demonstrate correctness of kPS, we generated $10^{4}$ paths starting
from state $(127,127)$ and ending at the absorbing boundary $\partial\mathbb{A}$
of the basin containing $N=2^{13}$ states, using both kPS and kMC
at T=2.5. The perfect match between the two estimated distributions
of exit times is shown in Fig.~\ref{fig2}.a. The mean times of exit
to $\partial\mathbb{A}$ are plotted as a function of the number of
eliminated states at T$=2.5$ and T=1.0 in Fig.~\ref{fig2}.b, while
the costs of both methods are compared in Fig.~\ref{fig2}.c. At
T=1.0, kMC trajectories are trapped and never reach $\partial\mathbb{A}$:
in this case we plot the expectation value for the number of kMC hops
required to exit $\mathbb{E}$ which is always available after path
factorization \cite{wales:2009}. We observe that the kPS cost scales
as $N^{3}$, as expected for this factorization, and exceeds that
of kMC for $N>2^{12}$ at T=2.5. However, at T=1 trapping becomes
severe rendering the standard kMC inefficient and the wall clock speedup
achieved by kPS is four orders of magnitude for $N=2^{15}$. We observe
that in kPS the net cost of generating an exit trajectory is nearly
independent of the temperature but grows exponentially with the decreasing
temperature in kMC. At the same time, an accurate measure of the relative
efficiency of kPS and kMC is always available in path factorization,
allowing one to revert to the standard kMC whenever it is relatively
more efficient. Thus, when performed correctly, a stochastic simulation
combining kPS and kMC should always be more efficient than kMC alone. 

As a second illustration, we apply kPS to simulate the kinetics of
copper precipitation in iron within a lattice model parameterized
using electronic structure calculations~\cite{soisson:2007}. The
simulation volume is a periodically replicated fragment of the body
centered cubic lattice with 128$\times$128$\times$128 sites on which
28,163 Cu atoms are initially randomly dispersed. Fe atoms occupy
all the remaining lattice sites except one that is left vacant allowing
atom diffusion to occur by vacancy (V) hopping to one of its NN lattice
sites. Its formation energy being substantially lower in Cu than in
Fe, the vacancy is readily trapped in Cu precipitates rendering kMC
grossly inefficient below $550$~K~\cite{soisson:2007}. Whenever
the vacancy is observed to attach to a Cu cluster, we perform kPS
over a pre-charted set $\mathbb{E}$ containing $N$ trapping states
that correspond to all possible vacancy positions inside the $\mathrm{VCu_{N-1}}$
cluster containing $N-1$ Cu atoms: the shape of the trapping cluster
is fixed at the instant when the vacancy first attaches. The fully
factored matrix $\mathbf{P}^{(N)}$ is then used to propagate the
vacancy to a lattice site just outside the fixed cluster shape which
is often followed by vacancy returning to the same cluster. If the
newly formed trapping cluster has the same shape as before, the factorized
matrix is used again to sample yet another escape. However a new path factorization (kPS cycle)
is performed whenever the vacancy re-attaches to the same Cu cluster
but in a different cluster shape or attaches to another Cu cluster
(see the Supplementary Material for additional simulation details~\cite{athenes:2014}). 

\begin{figure}[!t]
\includegraphics[width=0.45\textwidth]{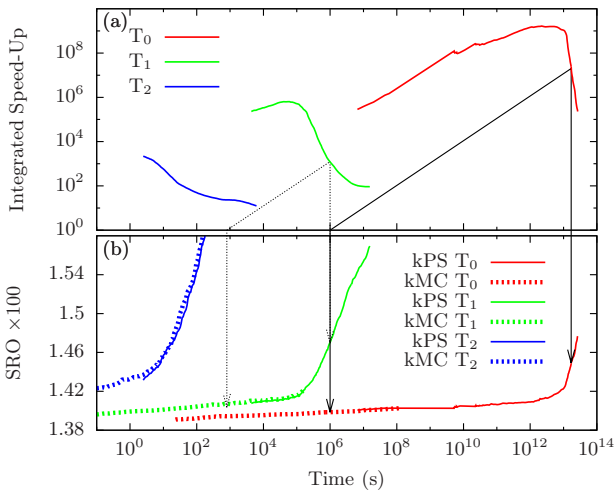} \caption{(a) Integrated speed-up plotted as a function of the physical time
simulated by kPS; (b) Time evolution of the averaged SRO in kPS and
kMC simulations at three different temperatures. }

\label{fig3} 
\end{figure}

We simulated copper precipitation in iron at three different temperatures
T$_{0}=273$~K, T$_{1}=373$~K and T$_{2}=473$~K for which the
atomic fraction of Cu atoms used in our simulations significantly
exceeds copper solubility limits in iron. Defined as the ratio of
physical time simulated by kPS to that reached in kMC simulations
over the same wall clock time, the integrated speed-up is plotted
in Fig.~\ref{fig3}.a. as a function of the physical time simulated
by kPS (averaged over 41 simulations for each method and at each temperature).

The precipitation kinetics are monitored through the evolution of
the volume-averaged Warren-Cowley short-range order (SRO) parameter~\cite{athenes:2000}
shown in Fig.~\ref{fig3}.b both for kPS and kMC simulations. At
T$_{0}$ and T$_{1}$ the kinetics proceed through a distinct incubation
stage reminescent to a time lag associated with repeated re-dissolution
of subcritical nuclei prior to reaching the critical size in the classical
nucleation theory~\cite{clouet:2010}. However, ``incubation''
observed here is of a distinctly different nature since all our simulated
solid solutions are thermodynamically unstable and even the smallest
of Cu clusters, once formed, never dissolve. At all three temperatures
the growth of $\mathrm{VCu_{N-1}}$ clusters is observed to proceed
not through the attachment of mobile V-Cu dimers but primarily through
the cluster's own diffusion and sweeping of neighboring immobile Cu
monomers~\cite{athenes:2014}. This is consistent with an earlier
study that also suggested that, rather counter-intuitively, the diffusivity
of $\mathrm{VCu_{N-1}}$ clusters should increase with the increasing
$N$ before tapering off at $N=30\div100$ (see Fig.~9 of Ref.~\onlinecite{soisson:2007}).
We futher observe that at T$_{0}$ the cross-over from the slow initial
``incubation'' to faster ``agglomeration'' growth seen on \ref{fig3}.b
occurs concomittantly with the largest cluster growing to 15-16 Cu
atoms~\cite{athenes:2014}. Individual realizations of the stochastic
precipitation kinetics reveal that, in addition to $N=15$, cluster growth
slows down again once the cluster reaches $N=23$, 27, 35 and so on (see
figure S4 in the Supplementary Materials). Leaving precise characterization
of these transitions to future work, we speculate that the observed 
``magic numbers'' correspond to compact clusters with fully filled 
nearest-neighbor shells in which vacancy trapping is particularly 
strong reducing the rate of shape-modifying vacancy escapes required 
for cluster diffusion. 

Numerically, as expected, the integrated speed-up rapidly increases 
with the decreasing temperature as vacancy trapping becomes more severe. 
Two line segments of unit slope and two pairs of vertical arrows are 
drawn in Fig.~\ref{fig3} to compare evolution stages achievable 
within kPS and kMC over the same wall clock time. As marked by the 
pair of two solid vertical arrows on the right, the integrated speed-up 
exceeds seven orders of magnitude at T$_{0}$. Subsequent reduction 
in the speed-up concides with the transition into the agglomeration 
regime where increasingly large VCu clusters repeatedly visit increasingly 
large number of distinct shapes~\cite{footnote:cluster}. Unquestionably, the 
efficiency of kPS simulations for this particular model can be improved
by indexing distinct cluster shapes for each cluster size and storing
the path factorizations to allow for their repeated use during the
simulations~\cite{beland:2011}. In any case, given its built-in
awareness of the relative cost measured in kMC hops, kPS is certain
to enable more efficient simulations of diffusive phase transformations
in various technologically important materials. In particular, it
is tempting to relate an anomalously long incubation stage observed
in aluminium alloys with Mg, Si and Se additions~\cite{pogatscher:2014}
to possible trapping of vacancies on Se, similar to the retarding
effect of Cu on the ageing kinetics reported here for the Fe-Cu alloys.

In summary, we developed a kinetic Path Sampling algorithm suitable
for simulating the evolution of systems prone to kinetic trapping.
Unlike most other algorithms dealing with this numerical bottleneck~\cite{novotny:1995,boulougouris:2007,barrio:2013,mason:2004,puchala:2010,beland:2011},
kPS does not require any \textit{a priori} knowledge of the properties
of the trapping basin. It relies on an iterative path factorization
of the evolution operator to chart possible escapes, measures its
own relative cost and reverts to standard kMC if the added efficiency
no longer offsets its computational overhead. At the same time, the
kPS algorithm is exact and samples stochastic trajectories from the
same statistical ensemble as the standard kMC algorithms. Being immune
to kinetic trapping, kPS is well positioned to extend the range of
applicability of stochastic simulations beyond their current limits.
Furthermore, kPS can be combined with spatial protection \cite{bulatov:2006}
and synchronous or asynchronous algorithms to enable efficient parallel
simulations of a still wider class of large-scale stochastic models
~\cite{shim:2005,merryck:2007,martinez:2011,jefferson:1989}. 
\begin{acknowledgments}
This work was supported by Defi NEEDS (Project MathDef) and Lawrence
Livermore National Laboratory's LDRD office (Project 09-ERD-005) and
utilized HPC resources from GENCI-{[}CCRT/CINES{]} (Grant x2013096973).
This work was performed under the auspices of the U.S. Department
of Energy by Lawrence Livermore National Laboratory under Contract
DE-AC52-07NA27344. The authors wish to express their gratitude to
T. Opplestrup, F. Soisson, E. Clouet, J.-L. Bocquet, G. Adjanor and
A. Donev for fruitful discussions. 
\end{acknowledgments}
\bibliographystyle{prl}

\begin{thebibliography}{10}
\bibitem{vankampen:2007} N. G. van Kampen, Stochastic Processes in
Physics and Chemistry (Elsevier Science, 2007).

\bibitem{redner:2001} S. Redner, A guide to first-passage processes,
Cambridge University Press (2001).

\bibitem{lanore:1974} J.-M. Lanore, Rad. Effects \textbf{22}, 153
(1974).

\bibitem{BKL} A. Bortz, M. Kalos, J. Lebowitz, J. Comp. Phys. \textbf{17},
10 (1975).

\bibitem{gillespie:1977} D. Gillespie, J. Chem. Phys. \textbf{81},
2340 (1977).

\bibitem{dellago:2002} P. Bolhuis, D. Chandler, C. Dellago and P.
Geissler, Ann. Rev. Phys. Chem. \textbf{53}, 291 (2002).

\bibitem{sun:2006} S. X. Sun, Phys. Rev. Lett. \textbf{96} 210602
(2006) and \textbf{97}, 178902 (2006).

\bibitem{harland:2007} B. Harland and X. Sun, J. Chem. Phys. \textbf{127}
104103 (2007).

\bibitem[9]{vansiclen:2007} C. D. Van Siclen, J. Phys.: Condens.
Matter 19, 072201 (2007).

\bibitem{eidelson:2012} N. Eidelson and B. Peters J. Chem. Phys.
\textbf{137}, 094106 (2012).

\bibitem{mora:2012} T. Mora, A. M. Walczak and F. Zamponi, Phys.
Rev. E \textbf{85}, 036710 (2012).

\bibitem{manhart:2013} M. Manhart and A. V. Morozov, Phys. Rev. Lett.
\textbf{111} 088102 (2013).

\bibitem{footnote:me} The evolution operator is obtained by integrating
the ME $\mathbf{\dot{v}}^{\mathrm{T}}(t)=\mathbf{v}^{\mathrm{T}}(t)\mathbf{M}$
from $t$ to $t+\tau$ and identifying the formal solution with $\mathbf{v}^{\mathrm{T}}(t+\tau)=\mathbf{v}^{\mathrm{T}}(t)\mathbf{P}(t,t+\tau)$
where $\mathbf{v}^{\mathrm{T}}(t)$ is the state-probability (row)
vector at $t$.

\bibitem{note:transition} If the system is in $\beta$ at a given
time, then the state-probability vector at a later time $\tau$ is
$\mathbf{v}_{\beta}^{\mathrm{T}}(\tau)=\mathbf{1}_{\beta}^{\mathrm{T}}\mathbf{P}(0,\tau)$
where $\mathbf{1}_{\beta}^{\mathrm{T}}$ denotes the row vector whose
$\beta$th entry is one and the other ones are zero. The entries of
$\mathbf{v}_{\beta}^{\mathrm{T}}(\tau)$ are $P_{\beta\gamma}(0,\tau)$
and can be used as transition probabilities for kMC moves from $\beta$.

\bibitem{novotny:1995} M. A. Novotny, Phys. Rev. Lett. { \textbf{74},
1 (1995).}

\bibitem{boulougouris:2007} G. Boulougouris and D. Theodorou, J.
Chem. Phys. \textbf{127}, 084903 (2007).

\bibitem{barrio:2013} M. Barrio, A. Leier and T. Marquez-Lago, J.
Chem Phys. \textbf{138}, 104114 (2013).

\bibitem{nandipati:2010} G. Nandipati, Y. Shim and J. G. Amar, Phys.
Rev. B \textbf{81} 235415 (2010).

\bibitem{moler:2003} C. Moler and C. Van Loan, SIAM Rev. \textbf{45},
3 (2003).

\bibitem{athenes:1997} M. Ath\`enes, P. Bellon and G. Martin, Phil.
Mag. A, \textbf{76}, 565 (1997).

\bibitem{trygubenko:2006} S. Trygubenko and D. Wales, J. Chem. Phys.
\textbf{124}, 234110 (2006).

\bibitem{wales:2009} D. Wales, J. Chem. Phys. \textbf{130}, 204111
(2009).

\bibitem{serebrinsky:2011} S. A. Serebrinsky, Phys. Rev. E \textbf{83},
037701 (2011).

\bibitem{athenes:2014} See Supplemental Material below for the connection between path factorization
and Gauss-Jordan elimination method and for a proof that the algorithm
is correct for $N=0$.

\bibitem{limoge:1990} Y. Limoge and J.-L. Bocquet, Phys. Rev. Lett.
\textbf{65 }, 60 (1990).

\bibitem{soisson:2007} F. Soisson, C. C. Fu, Phys. Rev. B \textbf{76},
214102 (2007).

\bibitem{athenes:2000} M. Ath\`enes, P. Bellon and G. Martin, Acta
Mat. \textbf{48}, 2675, (2000).

\bibitem{clouet:2010} E. Clouet, ASM Handbook Vol. 22A, Fundamentals
of Modeling for Metals Processing D. U. Furrer and S. L. Semiatin
(Eds.), pp. 203-219 (2010).

\bibitem{footnote:cluster} To understand the origin of the efficiency
decrease, we have monitored the number of distinct shapes of the vacancy-copper
cluster. For a V-Cu$_{30}$ cluster, we found that, over the $10^{3}$
last factorizations that have been performed, there are only 21 different
cluster shapes and that the 5 most frequent shapes occur with a frequency
of about 60\%.

\bibitem{beland:2011} L. K. B\'eland, P. Brommer, F. El-Mellouhi, J.
F. Joly and N. Mousseau, Phys. Rev. E \textbf{84}, 046704 (2011).

\bibitem{pogatscher:2014} S. Pogatscher, H. Antrekowitsch, M. Werinos,
F. Moszner, S. S. A. Gerstl, M. F. Francis, W. A. Curtin, J. F. L\"offler
and P. J. Uggowitzer, Phys. Rev. Lett. \textbf{112}, 225701 (2014).

\bibitem{mason:2004} D. Mason, R. Rudd and A. Sutton, Comp. Phys.
Comm. \textbf{160}, 140 (2004).

\bibitem{puchala:2010} B. Puchala, M. Falk and K. Garikipati, J.
Chem. Phys. \textbf{132}, 134104 (2010).

\bibitem{bulatov:2006} T. Opplestrup, V. V. Bulatov, G. H. Gilmer,
M. H. Kalos, and B. Sadigh, Phys. Rev. Lett. \textbf{97}, 230602 (2006).

\bibitem{shim:2005} Y. Shim and J. G. Amar, Phys. Rev. B \textbf{71},
115436 (2005).

\bibitem{merryck:2007} M. Merrick and K. Fichthorn, \textbf{75},
011606 (2007).

\bibitem{martinez:2011} E. Mart\'{i}nez and P. R. Monasterio, J.
Marian, J. Comput. Phys. \textbf{230}, 1359 (2011).

\bibitem[37]{jefferson:1989}F. Wieland, and D. Jefferson, Proc. 1989
Int'l Conf. Parallel Processing, Vol.III, F. Ris, and M. Kogge, Eds.,
pp. 255-258.\end{thebibliography}

\pagebreak
\widetext
\begin{center}
\textbf{\large Supplemental Materials: Path Factorization Approach to Stochastic Simulations}
\end{center}
\setcounter{equation}{0}
\setcounter{figure}{0}
\setcounter{table}{0}
\setcounter{page}{1}
\makeatletter
\renewcommand{\theequation}{S\arabic{equation}}
\renewcommand{\thefigure}{S\arabic{figure}}
\renewcommand{\bibnumfmt}[1]{[S#1]}
\renewcommand{\citenumfont}[1]{S#1}

\section{Path factorization with Gauss-Jordan elimination method}

Here, we describe the iterative procedure aiming at constructing the set of stochastic matrices $\left\{ \mathbf{P}^{(n)} \right\}_{0 \leq n \leq N}$. Let us define two matrices $\mathbf{U}$ and $\boldsymbol{\mathcal{L}}$
whose entries are initially set to zero. At the $n$th iteration,
the $n$th row of $\mathbf{U}$ and the $n$th column of $\boldsymbol{\mathcal{L}}$
are filled by setting $U_{n\gamma}={P}_{n\gamma}^{(n-1)}-I_{n\gamma}$
and $\mathcal{L}_{\beta n}=-{P}_{\beta n}^{(n-1)}/{U_{nn}}$, where
the pivot entry $U_{nn}=P_{nn}^{(n-1)}-1$ is the negative of the escape probability from state $n$ and $P_{nn}^{(n-1)}$ is the probability
of a round-trip from the same state. Given stochastic matrix $\mathbf{P}^{(n-1)}$,
the probability of an indirect transition from $\beta$ to $\gamma>n$
via $n$ is 
\begin{equation}
P_{\beta n}^{(n-1)}\sum_{f=0}^{+\infty}\Big[P_{nn}^{(n-1)}\Big]^{f}P_{n\gamma}^{(n-1)}=\mathcal{L}_{\beta n}U_{n\gamma},
\end{equation}
where the sum accounts for the probabilities of all possible round-trips
from $n$. Since $P_{\beta\gamma}^{(n-1)}=0$ for $\gamma\leq n-1$,
matrix $\mathbf{U}$ remains upper triangular after the $n$th row
addition and the equality $\mathcal{L}_{\beta n}U_{n\gamma}=-P_{\beta\gamma}^{(n-1)}$
remains satisfied for $\gamma\leq n$. The $n$th iteration is completed
by constructing $\mathbf{P}^{(n)}$ as follows: 
\begin{equation}
{P}_{\beta\gamma}^{(n)} = {P}_{\beta\gamma}^{(n-1)}+\mathcal{L}_{\beta n}U_{n\gamma},\label{eq:gauss-jordan}
\end{equation}
where $P_{\beta\gamma}^{(n)}=0$ for $\gamma\leq n$, as required. This property holds by induction on $n$ up to $n=N$.
After adding $\mathcal{L}_{\beta n}U_{n\gamma}$ in~\eqref{eq:gauss-jordan},
the probabilities of transitions from $\beta$ to $\gamma$ ($\gamma>n$)
subsume the cancelled probability of all possible transitions from
$\beta$ to $n$. This ensures that the transformed matrices $\mathbf{P}^{(n)}$
remain stochastic. 

The present path factorization~\eqref{eq:gauss-jordan} is connected to Gauss-Jordan elimination method. Summing $P^{(\nu)}_{\beta \gamma }- P^{(\nu-1)}_{\beta \gamma } = \mathcal{L}_{\beta \nu}  U_{\nu \gamma}$ from $\nu = \beta$ to $\nu=n$ when $\beta \leq n $ or from $\nu= n+1$ to $\nu= \beta -1$ when $n < \beta -1 $ yields the relation 
\begin{equation} \nonumber
 P^{(n)}_{\beta \gamma } = P^{(\beta-1)}_{\beta \gamma } + \sum_{\nu \geq \beta, ~\nu \leq n} \mathcal{L}_{\beta \nu}  U_{\nu \gamma} - \sum_{\nu > n,~\nu < \beta} \mathcal{L}_{\beta \nu} U_{\nu \gamma}. 
\end{equation}
Let $\mathbf{L}$ be the lower triangular matrix defined by $L_{\beta \gamma} = - \mathcal{L}_{\beta \gamma} $ for $\beta > \gamma $ and $L_{\beta \gamma} = I_{\beta \gamma}$ otherwise. Let also denote the upper triangular matrix $\mathbf{L}+\bm{\mathcal{L}}$  by $-\mathbf{U}^\#$. Then substituting $I_{\beta \gamma} + \sum_\nu I_{\beta \nu } U_{\nu \gamma} $ for $ P^{(\beta-1)}_{\beta \gamma }$ and identifying yields the equivalent relation ($\nu \geq 1$)
\begin{equation}\label{eq:incremental}
P^{(n)}_{\beta \gamma } = I_{\beta \gamma } -\sum_{\nu \leq n} U^\#_{\beta \nu } U_{\nu \gamma} + \sum_{\nu > n } L_{\beta \nu } U_{\nu \gamma }. 
\end{equation}
Choosing $n > \max (\beta,\gamma)$ in Eq.~\eqref{eq:incremental} entails ${P}^{(n)}_{\beta \gamma}=0$ and $(\mathbf{U}^\#\mathbf{U})_{\beta \gamma} = I_{\beta \gamma}$ , while setting $n=0$ yields $\mathbf{LU}=\mathbf{P}^{(0)}-\mathbf{I}=\tau \mathbf{M}$. Factorizations of $\mathbf{I}$ and $\tau\mathbf{M=P}^{(0)}-\mathbf{I}$ result from transformations above and below the eliminated pivot, respectively. This corresponds to the Gauss-Jordan pivot elimination method. 

The conditional probabilities used in the randomization procedure write 
\begin{equation}
R_{\beta\gamma}^{(n)}=\frac{P_{\beta\gamma}^{(n-1)}}{{P}_{\beta\gamma}^{(n)}}=1-\frac{\mathcal{L}_{\beta n}U_{n\gamma}}{{P}_{\beta\gamma}^{(n)}}\label{eq:S_ratio}. 
\end{equation}
By resorting to~\eqref{eq:gauss-jordan} and~\eqref{eq:S_ratio} and by using the stored entries of $\mathbf{U}$ and $\bm{\mathcal{L}}$, the information necessary for evaluating the conditional probabilities used in the iterative reverse randomization (See Sec.~\ref{algorithm} below) can be easily retrieved. 

\begin{figure}[!h]
\includegraphics[width=0.8\textwidth]{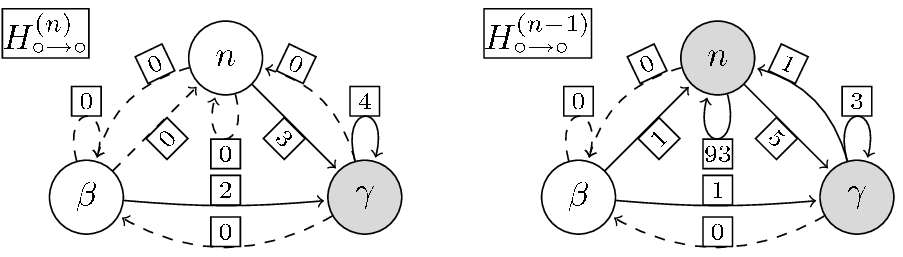}
\caption{Schematic diagram illustrating the random construction of $\mathbf{H}^{(n-1)}$ from $\mathbf{H}^{(n)}$ with $\beta < n < \gamma \le N$: dashed and solid arrows point to eliminated and non-eliminated states, respectively. Arrow labels indicate typical numbers of transitions. A transition out of an eliminated state means that an exiting path starts from this state: here, two paths start from $\beta$ and three from $n$.}
\label{fig:diagram}
\end{figure}

\section{Algorithm \label{algorithm}}

To show how the space-time randomization procedure is implemented in practice, the following preliminary definitions are required. The binomial law of trial number $h \in \mathbb{N}$ and success probability $r$ is denoted by $\mathrm{B}(h,r)$. The probability of $s$ successes is $\binom{h}{s} r^s (1-r)^{h-s}$. 
The negative binomial law of success number $h$ and success (escape) probability $1-p$ is denoted by  $\mathrm{NB} \left( h, 1-p \right)$. The probability of $f$ failures before the $h$-th success is $\binom{f+h-1}{f} p^f (1-p)^h$ where $p$ is the failure or flicker probability (flickers will correspond to round-trips from a given state).
The gamma law of shape parameter $h$ and time-scale $\tau$ is denoted by $\Gamma (h,\tau)$. 
$\mathrm{C}_\alpha$ denotes the categorical laws whose probability vector is the $\alpha$-th line of $\mathbf{P}^{(N)}$ if $\alpha \leq N$ or of the stochastic matrix obtained from $\mathbf{P}^{(0)}$ by eliminating the single state $\alpha > N$. The symbol $\sim$ means ``is a random variate distributed according to the law that follows''.
Let $\mathbb{A}$ and $\mathbb{T} = \overline{ \mathbb{A} \cup \mathbb{E}}$ denote the set of absorbing states and the set of noneliminated transient states, respectively. 
The absorbing boundary $\partial \mathbb{A}$ constains the state of  $\mathbb{A}$ that can be reached directly from  $\mathbb{E} \cup \mathbb{T}$. 
Let $\mathbf{h}$ be a vector and $\alpha$ denote the current state of the system.  

The cyclic structure of the algorithm, referred to as kinetic path sampling (KPS), is  
\begin{enumerate}[a.]
 \item compute $\mathbf{P}^{(N)}$ by iterating~\eqref{eq:gauss-jordan} on $n$ from $1$ to $N$, and  
 label the states connected to $\mathbb{E}$ in ascending order from $N+1$ to $N_c$ through appropriate permutations; define  $\mathbb{A}$ and $\mathbb{T}$ ; set the entries of $\mathbf{H}^{(N)}$ and $\mathbf{h}$ to zero; 
 \item \label{item:run} draw $\gamma \propto \mathrm{C}_\alpha$;  
increment $H_{\alpha,\gamma}^{(N)}$ or  $h_\alpha $ by one depending on whether $\alpha \in \mathbb{E}$ or $\alpha\in \mathbb{T}$; move current state $\alpha$ to $\gamma$ ; if new $\alpha \in \mathbb{A} $ goto (\ref{item:randomize}) otherwise repeat (\ref{item:run}); 
 
 \item \label{item:randomize} iterate in reverse order from $n=N$ to 1: 
    \begin{enumerate}[i.]
     \item  for $\beta \in \mathbb{E}\setminus \{ n \}$ and $\gamma \in \{n+1,..., N_c \}$ draw 
      \[ H^{(n-1)}_{\beta  \gamma} \sim \mathrm{B} \left(H^{(n)}_{\beta \gamma} , R^{(n)}_{\beta \gamma } \right);\]  
     \item for $\beta \in \mathbb{E}\setminus \{ n \}$ count the new hops from $\beta$ to $n$
      \[ H^{(n-1)}_{\beta n }  = \sum_{\gamma \in  \{n+1,..., N_c \} } H^{(n)}_{\beta \gamma} - H^{(n-1)}_{\beta \gamma}; \]
     \item for $\gamma \in \{n+1,..., N_c \}$ count the hops from $n$ to $\gamma$ 
      \[ H^{(n-1)}_{n \gamma}  = H_{n \gamma } ^{(n)} + \sum_{\beta \in \mathbb{E}\setminus \{ n \}} H^{(n)}_{\beta \gamma} - H^{(n-1)}_{\beta \gamma}; \] 
     \item  compute $h_n = \sum_{\gamma \in \{n+1,..., N_c \}} H^{(n-1)}_{n \gamma} $, the number of hops from $n$, and draw the flicker number 
         \[ H^{(n-1)}_{n n} \sim \mathrm{NB} \left( h_n , 1-P_{nn}^{(n-1)} \right) ; \]   \item store $T_n = h_n + H^{(n-1)}_{n n}$, the number of transitions from $n$, and deallocate $\mathbf{H}^{(n)}$, $\mathbf{P}^{(n)}$ and $\mathbf{R}^{(n)}$;
    \end{enumerate}  
 \item for $\ell \in \mathbb{T}$, store $T_\ell = h_\ell + f_\ell$ where $f _\ell \sim \mathrm{NB}(h_\ell,1-P_{\ell \ell}^{(0)})$; 
 \item evaluate $T^\mathbb{A} = \sum_{\ell \in \mathbb{E} \cup \mathbb{T}} T_\ell $,  the total number of flickers and hops associated with the path generated in~\eqref{item:run}; increment the physical time $t$ by $t^\mathbb{A} \sim \Gamma (T^\mathbb{A},\tau)$.  \label{item:gamma}
\end{enumerate}
After this cycle, the system has moved to the absorbing state $\alpha$ reached in~\eqref{item:run}. The gamma law $\Gamma (T^\mathbb{A},\tau)$ in~\eqref{item:gamma} simulates the time elapsed after performing $T^\mathbb{A}$ consecutive Poisson processes of rate $\tau^{-1}$. Indeed, after any hop or flicker performed with $\mathbf{P}^{(0)}$, the physical time must be incremented by a residence time  drawn in the exponential distribution of decaying rate $\tau^{-1}$. The way $\mathbb{E}$ and $\mathbb{A}$ are constructed at each cycle is specific to the application. 

In simulations of anomalous diffusion on a disordered substrate, $\mathbb{T} =\emptyset$ implying $\mathbb{A} = \overline{\mathbb{E}}$. 
Any transition from $\mathbb{E}$ reaches $\partial \mathbb{A}$. 
In constrast, when $\mathbb{T}$ is not empty, the generated path may return to set $\mathbb{E}$ several times prior to reaching $\partial \mathbb{A}$. With this more general set-up, several transitions exiting $\mathbb{E}$ are typically recorded in the hopping matrix, as illustrated in Fig.~\ref{fig:diagram}. This amounts to storing the path factorization and using it as many times as necessary. As a result, the elapsed physical time is generated for several escaping trajectories simultaneously.

\section{Disordered substrate model}

Let $\Omega(\beta)$ denote the subset of states $\nu$ whose distances to $\beta$ along (1,0) and (0,1)
directions are shorter than 48$a$, $a$ being the lattice parameter. 
Then, the energy landscape is constructed using 
\begin{eqnarray}
E_{\beta} & = & \epsilon\sum_{\nu\in\Omega(\beta)}(a_{\nu}+b_{\nu}) \\
E_{\beta\gamma}^{\mathrm{s}} & = & \epsilon\sum_{\nu\in\Omega(\beta)\cup\Omega(\gamma)}(a_{\nu}-b_{\nu}), 
\end{eqnarray}
where $a_{\nu}$ and $b_{\nu}$ are independent random variables taking
values -1 or 1 with equal probabilities.

\section{Simulation set-up for Fe-Cu system}

\begin{figure}[!b]
\includegraphics[width=0.4\textwidth]{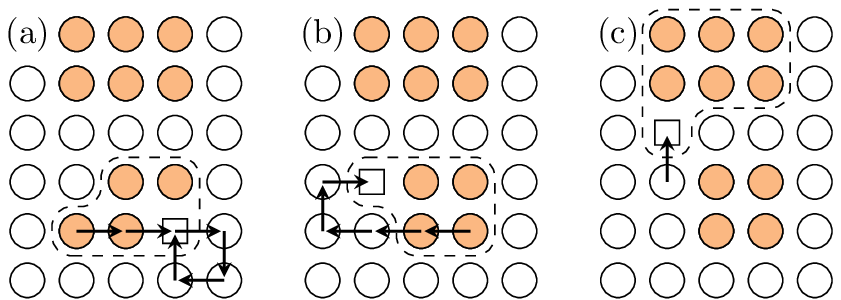}
\caption{Arrows indicate typical paths of the vacancy ($\square$) on a square lattice with Cu-like atoms colored in apricot. Dashed lines delimit the $V$Cu$_{N-1}$ cluster shapes. }
\label{figs2}
\end{figure}

In Fe-Cu, a KPS cycle starts by factoring the evolution operator when the vacancy binds to a Cu atom and forms a new $V$Cu cluster shape. Set $\mathbb{E}$ contains the configurations corresponding to the initial vacancy position and to the possible vacancy positions inside the neighboring Cu cluster (wherein the vacancy can exchange without moving any Fe atom). Cu and Fe atoms are unlabeled, hence the size $N$ of $\mathbb{E}$ is the number of Cu atoms in the Cu cluster plus one. 
Note that labeling the atoms would entail $N=(1+\mathrm{N_{Cu}})!$, making the simulations impractical for $\mathrm{N_{Cu}}>8$. 
The $n$th eliminated pivot in $\mathbb{E}$ corresponds to the least connected entry of $\mathbf{P}^{(n)}$. The stochastic matrix $\mathbf{P}^{(N)}$ is then used to evolve the system a first time and then each time the system returns to $\mathbb{E}$, as illustrated by the vacancy path of Fig.~\ref{figs2}.a. A KMC algorithm is implemented when the vacancy is embedded in the iron bulk. The set of non-eliminated transient states, $\mathbb{T}$, encompasses all states corresponding to the vacancy embedded in the iron bulk. 
In practice, the vacancy tends to return to the Cu cluster from which it just exited, meaning that the factorization is used many times during a KPS cycle, up to an average of 60 for a cluster of size 40 at $T_0=273$~K. 
One stops generating the path whenever the vacancy returns to the initial Cu cluster but with a different $V$Cu$_{N-1}$ cluster shape (see Fig.~\ref{figs2}.b) or reaches another Cu cluster (see Fig.~\ref{figs2}.c). The possible ending states define the absorbing boundary $\partial \mathbb{A}$. Then, KPS completes its cycle with space-time randomization. 

\begin{figure}[!t]
\includegraphics[width=0.45\textwidth]{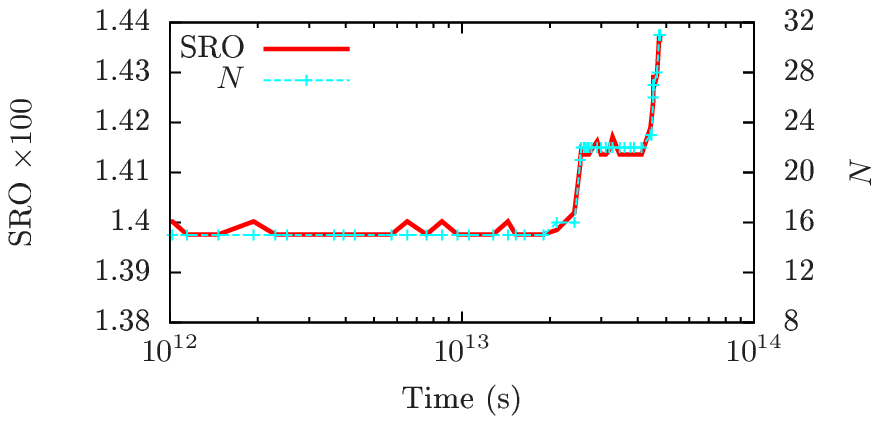}
\caption{ Concomittant evolution of the SRO parameter and of the size $N$ of the migrating Cu$_{N-1}$V cluster: single (non-averaged) ageing kinetics at $T_0=373$ K.}
\label{figs3}
\end{figure}

\begin{figure}[!h]
\includegraphics[width=0.45\textwidth]{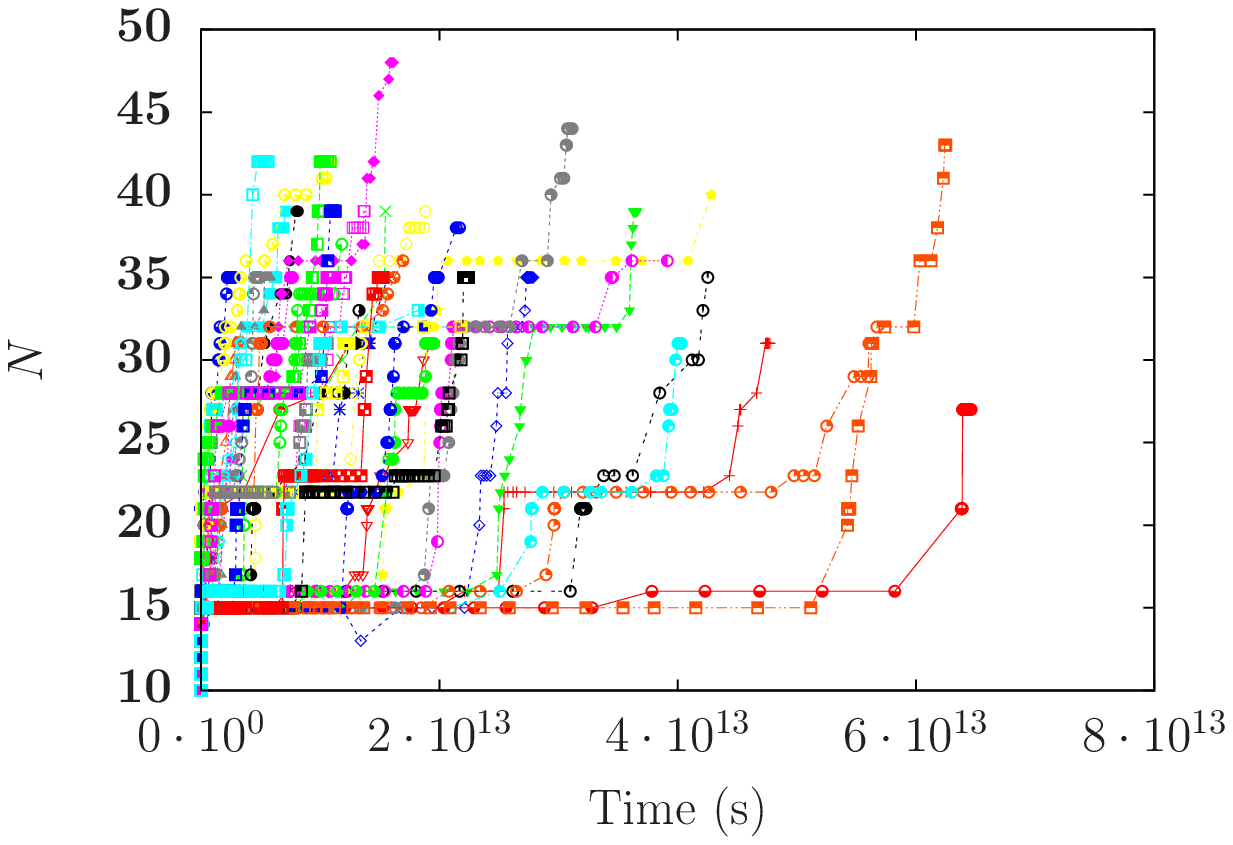}
\caption{Time evolution of the size $N$ associated with the migrating Cu$_{N-1}$V clusters for the 41 trajectories simulated at $T_0=373$ K.}
\label{figs4}
\end{figure}

The copper solubility limits in iron for the three simulated temperatures ($T_0=273$~K, $T_1=373$~K and $T_2=473$~K) are $4.39\times 10^{-5}$, $4.27\times 10^{-6}$ and $1.19\times 10^{-7}$, respectively. These limits are very small compared with $1.34 \times 10^{-2}$, the Cu concentration in the simulations. Supersaturations in copper are thus very important, meaning that the solid solution is unstable at these temperatures and that the Cu clusters that form during subsequent ageing unlikely dissolve during the initial incubation at $T_{0}$. We in fact observe that the cross-over from the slow initial ``incubation'' to faster growth occurs concomittantly
with a vacancy-containing copper cluster growing to 15-16 Cu atoms, as shown in Fig.~\ref{figs3}. Additional ``magic numbers'' around  $N=23$, 27, 35 are evidenced at $T_{0}$ by visualizing the 41 simulated trajectories displayed in Fig.~\ref{figs4}. These numbers correspond to the numbers of site in compact clusters with fully filled
nearest-neighbor shells.  

\section{Proof}

We now prove that the algorithm generates the correct distribution of first-passage times to $\mathbb{A}$ for $N=0$, i.e $\mathbb{E} = \emptyset$. 
In step~(e) of the KPS algorithm, we resort to the distributivity of the gamma law with respect to its shape parameter and decompose the first-passage time as follows: 
\begin{equation}
t^\mathbb{A} \sim  \sum_{\ell \in \mathbb{T}}  \Gamma (h_\ell + f_\ell,\tau)  \label{eq:gamma}, 
\end{equation}
where $h_\ell$ and $f_\ell$ are the generated numbers of hops and flickers from $n$. 
For any state $\ell \in \mathbb{T}$ that is visited $h$ times, the probability of having $f$ flickers before the $h$-th escape from $\ell$ is $\binom{h-1+f}{f} (P_{\ell \ell})^f (1-P_{\ell \ell})^h$, which corresponds to the probability mass of the negative binomial law of success number $h$ and success probability $1-P_{\ell \ell}$. In the following, we introduce the effective rate 
\begin{equation}
 k_\ell = \tau^{-1}(1-P_{\ell \ell}) = - M_{\ell \ell}. 
\end{equation}
The residence time associated with $h+f$ hops and flickers from state $\ell$ is distributed according to $\Gamma(h+f,\tau)$, the Gamma law of shape parameter $h+f$ and time-scale $\tau$. The probability mass of $\Gamma(h+f,\tau)$ law at $t$ being $\tau^{-1}\frac{1}{(h-1+f)!}(t/\tau)^{h-1+f} \exp[-t/\tau ]$, the overall probability to draw residence time $t$ after $h$ visits of state $\ell$  is obtained by summing the compound probabilities associated with all possible occurrences of the number $f$ of flickers as follows 
\begin{eqnarray}
\frac{1}{(h-1)!} \left( \frac{\tau k_\ell}{\tau} \right)^h t ^{h-1}  \sum_{f=0}^{+ \infty} \frac{1}{f!} \left[ \frac{t}{\tau} P_{\ell \ell} \right]^{f}  \exp \left[-\frac{t}{ \tau} \right] = \nonumber  \frac{ (k_\ell t )^{h-1} }{(h-1)!} k_\ell \exp\left[ - k_\ell t  \right].   \nonumber
\end{eqnarray}
Remarkably, the summation removes the dependence on $\tau$ and yields the distribution of the Gamma law of shape parameter $h$ and time-scale $1/k_\ell$, which corresponds to the convolution of $h$ decaying exponentials of rate $k_\ell$.  This is the expected distribution for the time elapsed after $h$ consecutive Poisson processes of rate $k_\ell$. Note that the standard KMC algorithm simply draws an escape time according to $\Gamma(1,1/k_\ell)$, the exponential law of rate $k_\ell$ for each visit of $\ell$, which is statistically equivalent. This amounts to prescribing a success probability of 1, which results in $\tau = 1/k_\ell$ and no failures. The proof that the algorithm is correct for $N \ge 1$ follows by induction on $N$ resorting to the iterative structure of the algorithm. 

Finally, it is worth noticing that trajectories escaping a trapping basin can in principle be analysed to provide information about the kinetic pathways since the numbers of transitions involving each eliminated state are randomly generated. Obtaining information about time correlation functions and commitor probabilities would require testing the involved Bernoulli trials sequentially instead of drawing a binomial or negative binomial deviates given a number of transitions, in order to construct the path. However, the computational cost of these additional operations scales linearly with the mean first-passage time for escaping a trapping basin.

\end{document}